\documentclass[pre,twocolumn,groupedaddress,showkeys,showpacs]{revtex4}

\usepackage{graphicx}
\usepackage{CJK}
\begin{document}
\begin{CJK*}{GBK}{}


\title{Scaling analysis of negative differential thermal resistance}


\author{Ho-Kei Chan $^{1}$}
\email{epkeiyeah@yahoo.com.hk}
\author{Dahai He $^{2}$}
\email{dhe@xmu.edu.cn}
\author{Bambi Hu $^{3}$,}


\affiliation{$^{1}$Department of Physical and Theoretical Chemistry, School of Chemistry, University of Nottingham, Nottingham NG7 2RD, U. K.}
\affiliation{$^{2}$Department of Physics, Xiamen University, Xiamen 361005, China.}
\affiliation{$^{3}$Department of Physics, University of Houston, Houston, Texas 77204-5005, U. S. A.}


\date{\today}
\begin{abstract}
\indent
Negative differential thermal resistance (NDTR) can be generated for any one-dimensional heat flow with a temperature-dependent thermal conductivity. In a system-independent scaling analysis, the general condition for the occurrence of NDTR is found to be an inequality with three scaling exponents: $n_{1}n_{2}<-(1+n_{3})$, where $n_{1}\in(-\infty,+\infty)$ describes a particular way of varying the temperature difference, and $n_{2}$ and $n_{3}$ describe, respectively, the dependence of the thermal conductivity on an average temperature and on the temperature difference. For cases with a temperature-dependent thermal conductivity, i.e. $n_{2}\neq0$, NDTR can \emph{always} be generated with a suitable choice of $n_{1}$ such that this inequality is satisfied. The results explain the illusory absence of a NDTR regime in certain lattices and predict new ways of generating NDTR, where such predictions have been verified numerically. The analysis will provide insights for a designing of thermal devices, and for a manipulation of heat flow in experimental systems, such as nanotubes.
\end{abstract}




\maketitle
\end{CJK*}
\section {Introduction}
\indent
The mechanism of negative differential thermal resistance (NDTR), i.e. the counterintuitive phenomenon of decreasing heat flux for increasing temperature difference, is a fundamental problem in modern thermal transport research \cite{R04}. NDTR plays a pivotal role in the operation of various model thermal devices, such as the thermal transistor \cite{R01}, the thermal-memory device \cite{R02}, the heat-current limiter \cite{R03} and the constant heat-current source \cite{R03}, and it also occurs in the thermal rectifier \cite{Old03,Old04}. Being central to the relatively new concept of phonon computing \cite{R04,R05}, NDTR is reminiscent of the phenomenon of negative differential electrical resistance in the operation of the tunnel diode \cite{R06} and of some other electrical devices \cite{R07}, where there exists a regime of decreasing electrical current for increasing voltage. In light of the importance of NDTR in the designing of thermal devices, and also of a general interest in the experimental manipulation of heat flow (e.g. nanotube-based thermal rectifiers \cite{R18,R19}, thermal cloaks \cite{Extra1,Extra2}), there has been much interest in understanding how NDTR can be systematically generated. In particular, studies have been carried out for various nonlinear lattices \cite{R08,R09,R10,R11,R12,R13}, namely the Frenkel-Kontorova (FK) model \cite{R08,R09,R10}, the $\phi^4$ model \cite{R10} and the Fermi-Pasta-Ulam (FPU) model \cite{R10,R11}. Each of these models is described by a Hamiltonian of the form
\begin{equation}\label{Eq_Intro}
H=\sum\limits_{i=1}^n \left[\frac{p_{i}^2}{2} + U(x_{i})\right] + \sum\limits_{i=1}^{n-1} V(x_{i+1}-x_{i})
\end{equation}
where $N$ is the total number of particles (system size) and, for the $i^{th}$ particle, $x_{i}$ is the position, $p_{i}$ is the instantaneous momentum, $U(x_{i})$ is the onsite potential, and $V(x_{i+1}-x_{i})$ is the potential of nearest-neighbour interaction. As in previous work \cite{R10}, the temperature $T_{i}$ of the $i^{th}$ particle is defined as proportional to the particle's kinetic energy:
\begin{equation}\label{Eq_Temp}
T_{i}\equiv \frac{\left<p_{i}^2\right>}{m_{i}} = m_{i}\left<v_{i}^2\right>
\end{equation}
where the chevrons denote a time average, $m_{i}=1$ is the particle's mass, and $v_{i}=p_{i}/m_{i}$ is the particle's speed. Eq. (\ref{Eq_Temp}) follows directly from the equipartition of energy, albeit the Boltzmann constant $k_{B}$ is absorbed into $T_{i}$. The dynamics of each lattice is modelled via non-equilibrium molecular dynamics simulations, and the boundary temperatures $T_{+}$ and $T_{-}$ at the ends of a chain are controlled using Langevin heat baths \cite{R14}. The equations of motion are solved using the velocity Verlet algorithm. With a rich variety in their nonlinear dynamics \cite{R14,R15}, these models have been used as building blocks for the devices mentioned above.

Recent studies \cite{R08,R10} suggest that the existence of an observable regime of NDTR is conditional upon the presence of a nonlinear onsite potential as well as a small system size, the latter being attributed to boundary effects, for example thermal boundary resistance \cite{R10}. For the various lattice models mentioned above, conditions for the occurrence of NDTR have so far been studied only in terms of specific system parameters, for example in terms of the onsite potential, the system size, and the spring constant, often with the additional constraint that the temperature is kept invariant at either end of an one-dimensional lattice. From the experimental point of view, however, it is not always easy to relate specific parameters of a lattice model with experimentally measurable parameters, and therefore studies of this kind might be of limited utility. On the other hand, from both the theoretical and experimental points of view, a fundamental understanding of the nature of NDTR, and a systematic method of manipulating the occurrence of NDTR, are both lacking. In this paper, a general theoretical analysis of NDTR is presented in terms of the temperature dependance of the experimentally measurable thermal conductivity, where this analysis would be useful for both experimentalists and theorists: It is shown that NDTR can be generated for \emph{any} one-dimensional heat flow with a temperature-dependent thermal conductivity and that the abovementioned additional constraint is\emph{ not} necessary for generating NDTR. The conclusions are applicable not only to specific theoretical models but also generally to heat flow in experimental systems.

\section {Theoretical analysis}

\begin{figure}[htbp]
\begin{center}\includegraphics[width=7.0125cm,height=5.41875cm]{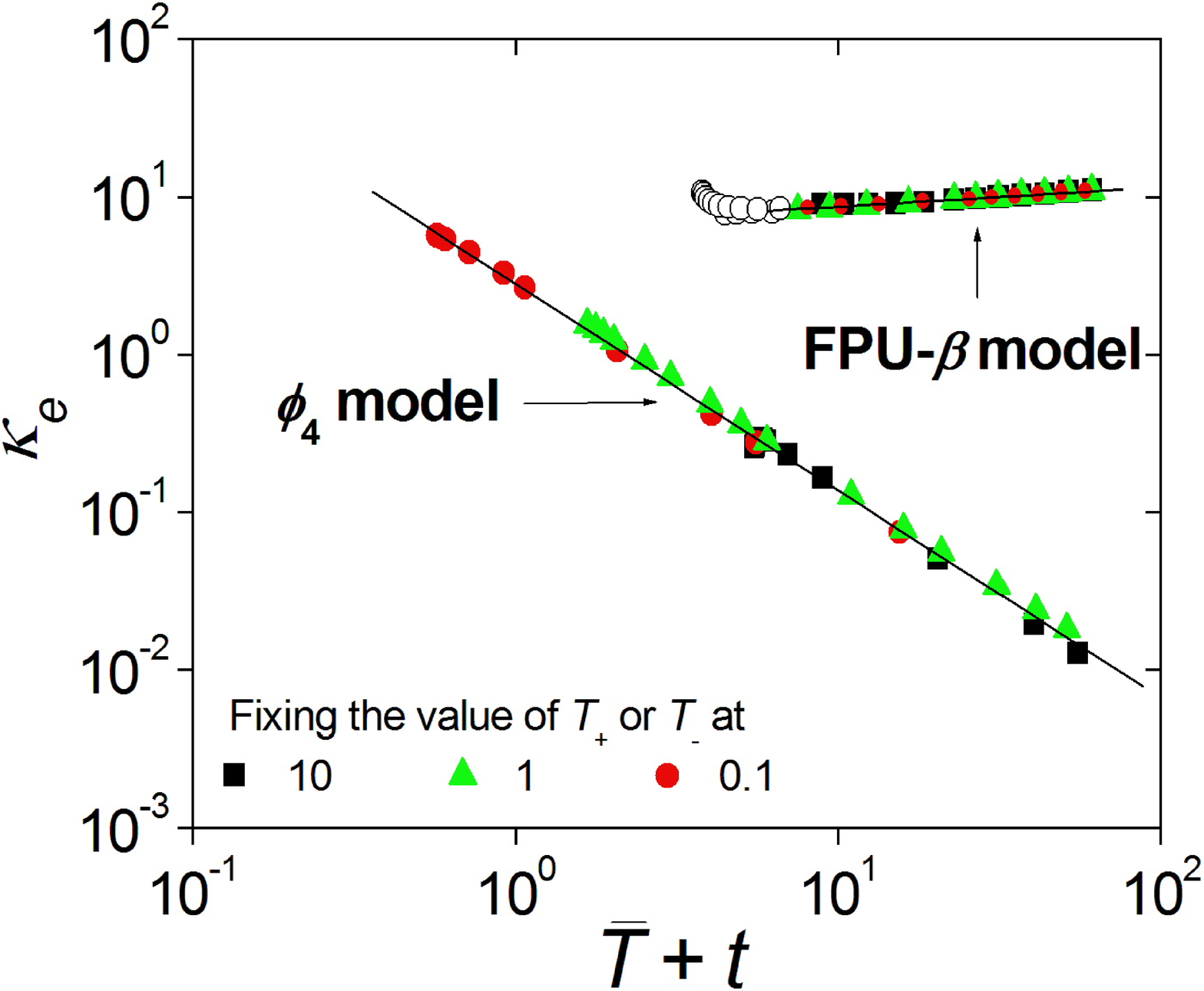}
\caption{Log-log plot of the effective thermal conductivity $\kappa_{e}$ against $\overline{T}+t$, where $\overline{T}\equiv (T_{+}+T_{-})/2$ is an average of the boundary temperatures. $t$ is a fitted parameter, with (a) $t\approx0.52$ for a $\phi^4$ chain of $N=32$ particles and (b) $t\approx3.73$ for a FPU-$\beta$ chain of $N=64$ particles. For both models, the data fit well to the power-law described by Eq. (\ref{Eq_1}) (within a specific range of $\overline{T}$ for the FPU-$\beta$ model), with a fitted slope of (a) $-\gamma\approx-1.313\pm0.009$ for the $\phi^4$ model and (b) $-\gamma\approx0.130\pm0.002$ for the FPU-$\beta$ model (The 'rings' denote data that deviate from the corresponding power law).
}\label{Fig_1}
\end{center}
\end{figure}

It has recently been shown that the condition for the occurrence of NDTR in the $\phi^4$ model (with an onsite potential of $\lambda x_{i}^{4}/4$, where $\lambda=1$ is the strength of the potential) can be described in terms of a scaling exponent $\gamma>0$ that depicts the temperature dependence of the system's thermal conductivity \cite{R10}: In this nonlinear lattice, the effective thermal conductivity $\kappa_{e}$ approximately follows a scaling temperature dependence (Fig. \ref{Fig_1}):
\begin{equation}\label{Eq_1}
\kappa_{e}(\overline{T})\equiv\frac{jN}{\Delta T}\approx C(\overline{T}+t)^{-\gamma}
\end{equation}
where $N$ is the total number of particles), $j$ is the heat flux, $\overline{T}\equiv (T_{+}+T_{-})/2$  is an average of the boundary temperatures $T_{+}>0$ and $T_{-}>0$, and $\Delta T\equiv T_{+}-T_{-}>0$ is the temperature difference across the lattice chain. $C>0$ and $t>0$ are some constants. As shown in Fig. \ref{Fig_1}, Eq. (\ref{Eq_1}) also holds for the FPU-$\beta$ model (with a $\beta(x_{i+1}-x_{i})^{4}/4$ in its potential of nearest-neighbour interaction, where $\beta=1$ is the strength of such nonlinear interaction) with $\gamma<0$ and $t>0$ within a specific range of $\overline{T}$. For the conventional cases of having the value of either $T_{+}$ or $T_{-}$ fixed \cite{R08,R09,R10,R11,R12,R13}, NDTR corresponds to
\begin{equation}\label{Eq_2}
{\left[\frac{\partial j(T_{\pm},\Delta T)}{\partial\Delta T}\right]_{T_{\pm}}}=\mp{\left[\frac{\partial j(T_{+},T_{-})}{\partial T_{\mp}}\right]_{T_{\pm}}} <0
\end{equation}
which implies
\begin{equation}\label{Eq_3}
\Delta T>\Delta T^{*}\equiv\frac{2(T_{\pm}+t)}{[\mp(\gamma-1)]}
\end{equation}
for
\begin{equation}\label{Eq_4}
\mp(\gamma-1)>0.
\end{equation}
For $\pm(\gamma-1)>0$, Eq. (\ref{Eq_2}) yields a disallowed condition of $\Delta T$ being smaller than a negative value. For cases with the value of $T_{-}$ being fixed, it simply follows from Eq. (\ref{Eq_4}) that $\gamma>1$ is the condition for the occurrence of NDTR. And for cases with the value of $T_{+}$ being fixed, the additional constraint $\Delta T^{*}<\Delta T_{max}=T_{+}$ implies $\gamma<1-2[1+(t/T_{+})]$; that is, $\gamma$ must be negative. It follows that, for the $\phi^{4}$ model with $\gamma>0$, cases of NDTR with the value of $T_{+}$ being fixed cannot occur, which is consistent with existing numerical findings \cite{R10}.

\begin{figure}[htbp]
\begin{center}\includegraphics[width=5.2cm,height=5.2cm]{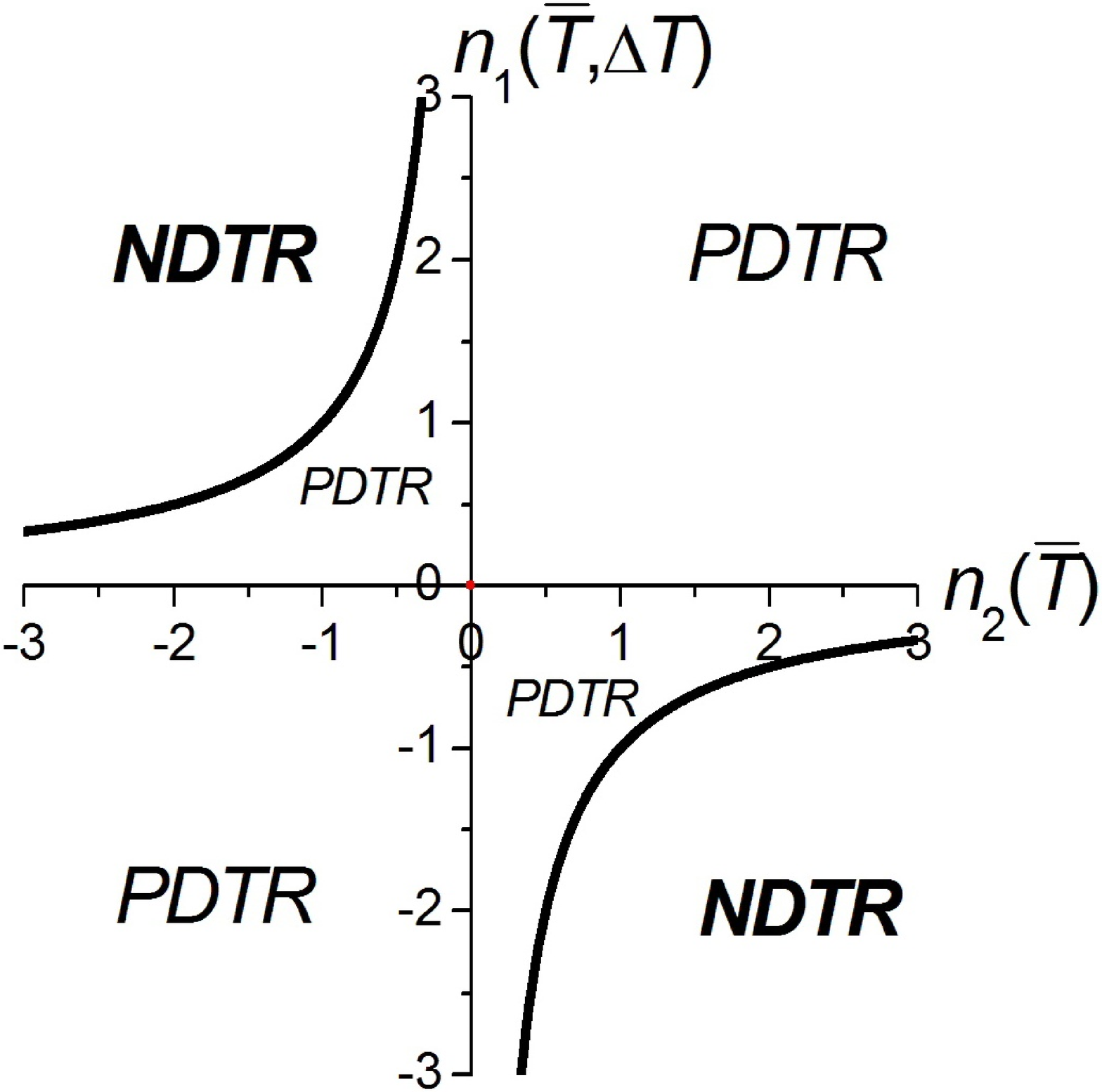}
\caption{A 'phase diagram' for Eq. (\ref{Eq_19}) [i.e. $\kappa_{e}\approx\kappa_{e}(\overline{T})$, $n_{3}(\overline{T},\Delta T)\approx 0$ and $n_{2}\approx n_{2}(\overline{T})$], illustrating at what combinations of $n_{1}(\overline{T},\Delta T)$ and $n_{2}(\overline{T})$ will NDTR occur. The acronym PDTR stands for the opposite effect of positive differential thermal resistance.
}\label{Fig_2}
\end{center}
\end{figure}

Despite its specific applicability to the $\phi^4$ model, the above theoretical analysis can readily be generalized to \emph{any} one-dimensional heat flow, not only in theoretical models but also in experimental systems. This can be achieved by removing any model- or system-specific features, e.g. the specific scaling temperature dependence of the effective thermal conductivity $\kappa_{e}$ as described by Eq. (\ref{Eq_1}). Starting from the general form
\begin{equation}\label{Eq_5}
\kappa_{e}=\kappa_{e}(\overline{T},\Delta T)=\kappa_{e}(T_{+},T_{-}),
\end{equation}
which describes the general dependence of this effective thermal conductivity on two externally controlled parameters, one can write
\begin{equation}\label{Eq_6}
j=j(\overline{T},\Delta T)=\frac{\kappa_{e}(\overline{T},\Delta T)\Delta T}{N}
\end{equation}
and
\begin{eqnarray}\label{Eq_7}
\lefteqn{dj(\overline{T},\Delta T)}\nonumber\\
&=&{{\left[\frac{\partial j(\overline{T},\Delta T)}{\partial \overline{T}}\right]}_{\Delta T}d\overline{T}}
+{{\left[\frac{\partial j(\overline{T},\Delta T)}{\partial \Delta T}\right]}_{\overline{T}}d\Delta T}\nonumber\\
\end{eqnarray}
Eq. (\ref{Eq_5}) generally takes into account any possible nonlinear response \cite{R10}, particularly at large values of $\Delta T$ where the thermal conductivity could be more than a temperature-dependent material property; it could depend not only on the system's average temperature but also on the temperature difference. This is analogous to the possible existence of a field-dependent electrical conductivity at large electric fields \cite{R16}, for example in the case of space-charge-limited conduction \cite{R17} where a power-law relation between the current density $j$ and the applied voltage $V$ exists. Using Eq. (\ref{Eq_6}), Eq. (\ref{Eq_7}) can be rewritten as
\begin{eqnarray}\label{Eq_8}
\lefteqn{dj(\overline{T},\Delta T)={\frac{\kappa_{e}(\overline{T},\Delta T)}{N}}\times}\nonumber\\
& &\left\{ [n_{2}(\overline{T},\Delta T) \Delta T] d\ln\overline{T}+[1+n_{3}(\overline{T},\Delta T)]d \Delta T\right\}
\end{eqnarray}
where the scaling exponents
\begin{equation}\label{Eq_9}
n_{2}(\overline{T},\Delta T)\equiv\left[\frac{\partial \ln \left[\kappa_{e}(\overline{T},\Delta T)\right]}{\partial \ln\overline{T}}\right]_{\Delta T}
\end{equation}
and
\begin{equation}\label{Eq_10}
n_{3}(\overline{T},\Delta T)\equiv\left[\frac{\partial \ln \left[\kappa_{e}(\overline{T},\Delta T)\right]}{\partial
\ln\Delta T}\right]_{\overline{T}}
\end{equation}
are hereafter referred to as the conductivity exponents; They describe, respectively, the dependence of the effective thermal conductivity $\kappa_{e}$ on $\overline{T}$ and $\Delta T$. On the other hand, the constraint of fixing the value of either $T_{+}$ or $T_{-}$ belongs to a generalized set of constraints, each of which can be described in terms of a particular curve
\begin{equation}\label{Eq_11}
F(\overline{T},\Delta T)=0
\end{equation}
in the $(\overline{T},\Delta T)$ parameter space. It follows that
\begin{eqnarray}\label{Eq_12}
\lefteqn{dF(\overline{T},\Delta T)=0}\nonumber\\
&=&{\left[\frac{\partial F(\overline{T},\Delta T)}{\partial \overline{T}}\right]_{\Delta T}d\overline{T}}
+{\left[\frac{\partial F(\overline{T},\Delta T)}{\partial \Delta T}\right]_{\overline{T}}d\Delta T}\nonumber\\
\end{eqnarray}
for the function $F(\overline{T},\Delta T)$, so that a corresponding scaling exponent
\begin{eqnarray}\label{Eq_13}
n_{1}(\overline{T},\Delta T)
\equiv-\frac{\left[{\frac{\partial F(\overline{T},\Delta T)}{\partial \ln\Delta T}}\right]_{\overline{T}}}
{\left[{\frac{\partial F(\overline{T},\Delta T)}{\partial \ln\overline{T}}}\right]_{\Delta T}}
=\left[\frac{d \ln\overline{T}}{d \ln \Delta T}\right]}_{F(\overline{T},\Delta T)=0
\end{eqnarray}
with respect to the constraint can be defined. The conventional cases of having the value of either $T_{+}$ or $T_{-}$ fixed correspond to particular linear relations between $\overline{T}$ and $\Delta T$ in the parameter space:
\begin{equation}\label{Eq_14}
dT_{\pm}=0,
\end{equation}
\begin{equation}\label{Eq_15}
\overline{T}=T_{\pm}\mp\frac{\Delta T}{2}=const\mp\frac{\Delta T}{2}
\end{equation}
and therefore
\begin{equation}\label{Eq_16}
n_{1}(\overline{T},\Delta T)=\mp\frac{\Delta T}{2\overline{T}}\in[0,\mp1].
\end{equation}
In the more general theoretical framework described above, NDTR corresponds to
\begin{eqnarray}\label{Eq_17}
\left[{\frac{\partial j(\overline{T},\Delta T)}{\partial \Delta T}}\right]_{F(\overline{T},\Delta T)=0}<0
\end{eqnarray}
which implies
\begin{equation}\label{Eq_18}
n_{1}(\overline{T},\Delta T)n_{2}(\overline{T},\Delta T)<-[1+n_{3}(\overline{T},\Delta T)]
\end{equation}
according to Eq. (\ref{Eq_8}). As long as $n_{2}(\overline{T},\Delta T)\neq0$, one can always ensure that this inequality is satisfied, by choosing a suitable value of $n_{1}(\overline{T},\Delta T)\in(-\infty,+\infty)$ where this value represents a particular way of varying the temperature difference $\Delta T$ across the system to generate NDTR. If $T_{+}$ and $T_{-}$ undergo almost the same infinitesimal increase or decrease, $\Delta T$ would have undergone almost no change as compared to the corresponding increase or decrease in $\overline{T}$, which implies $n_{1}\rightarrow\pm\infty$. It follows that any finite value of $n_{1}(\overline{T},\Delta T)\in(-\infty,+\infty)$ can be generated, so that Eq. (\ref{Eq_18}) can always be satisfied. Note that a more general condition for the occurrence of NDTR can be derived if one takes into account possible effects of thermal expansion, i.e. dependence of $N$ on $\overline{T}$, but Eq. (\ref{Eq_18}) is sufficient for providing insights into the fundamental nature of NDTR and for understanding a wide range of numerical results concerning the presence or illusory absence of a NDTR regime. Eq. (\ref{Eq_18}) states that NDTR is an artificial effect, depending on how the externally controlled temperature difference is varied. For cases with a negligible dependence of $\kappa_{e}$ on $\Delta T$, such as the examples in Fig. \ref{Fig_1}, Eq. (\ref{Eq_18}) can be simplified to
\begin{equation}\label{Eq_19}
n_{1}(\overline{T},\Delta T)n_{2}(\overline{T})<-1
\end{equation}
which can be depicted as a 'phase diagram' of the exponents $n_{1}(\overline{T},\Delta T)$ and $n_{2}(\overline{T})$ (Fig. \ref{Fig_2}). In this 'phase diagram', regions of NDTR and PDTR (positive differential thermal resistance) are separated by the curves of $n_{1}(\overline{T},\Delta T)n_{2}(\overline{T})=-1$. In the following, Eq. (\ref{Eq_1}), which describes a scaling temperature dependence of the effective thermal conductivity, is used as an example to illustrate the presence of a great variety of methods for generating NDTR in addition to the conventional approaches of fixing the value of either $T_{+}$ or $T_{-}$.

\section {Examples}

\begin{figure}[htbp]
\begin{center}\includegraphics[width=7.0125cm,height=5.41875cm]{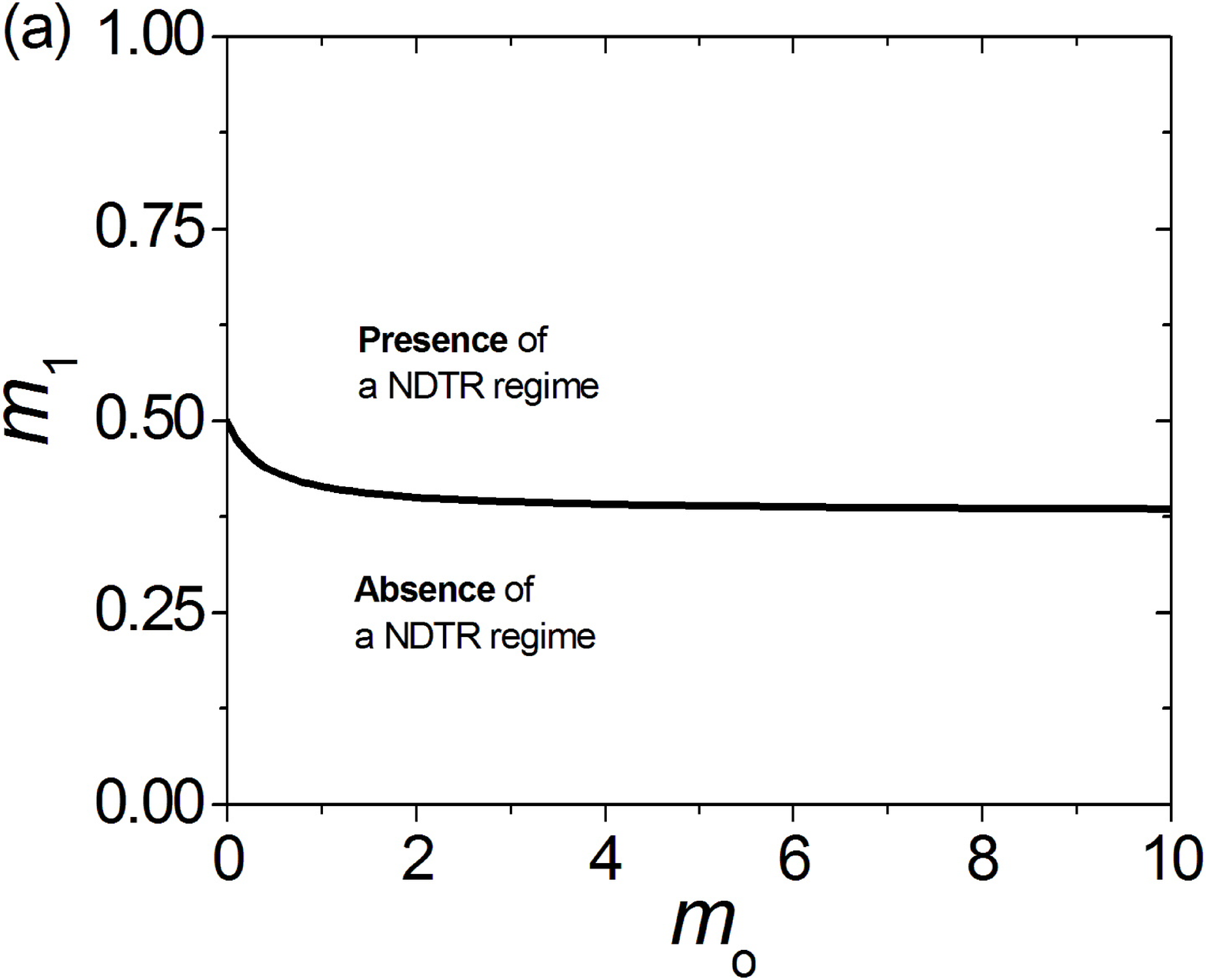}
\includegraphics[width=7.0125cm,height=5.41875cm]{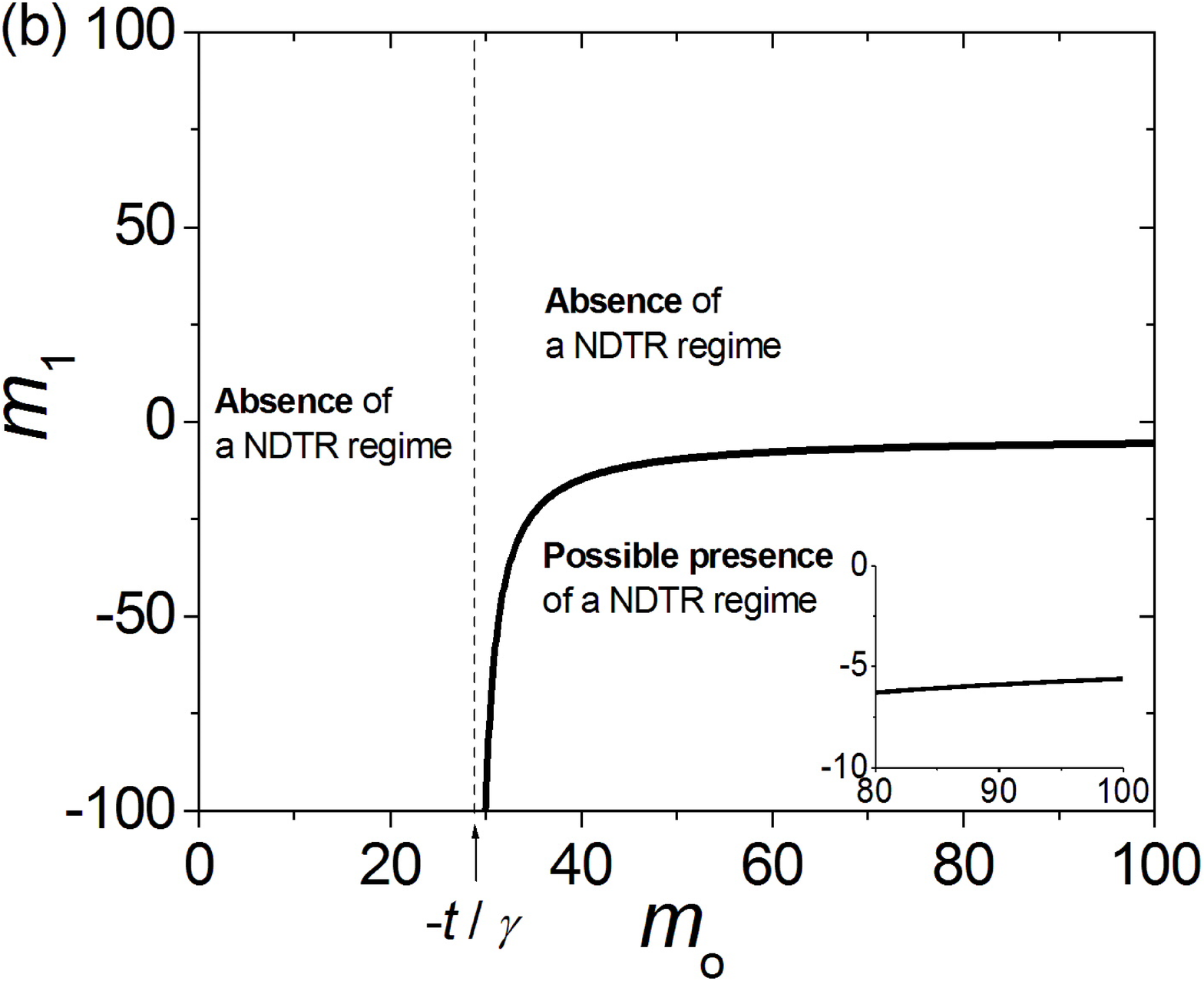}
\caption{'Phase diagrams' of $m_{1}$ and $m_{o}$ for (a) a $\phi^4$ chain of $N=32$ particles [$\gamma\approx1.313\pm0.009$ and $t \approx 0.52$] and (b) a FPU-$\beta$ chain of $N=64$ particles [$\gamma\approx-0.130\pm0.002$ and $t \approx 3.73$]. (a) For NDTR to occur in the $\phi^4$ model, $m_{1}$ must be greater than a positive value given by Eq. (\ref{Eq_29}) (solid line). (b) For NDTR to occur in the FPU-$\beta$ model, $m_{1}$ must be smaller than a negative value given by Eq. (\ref{Eq_30}) (solid line), and $m_{o}$ must be greater than $-t/\gamma$ according to Eq. (\ref{Eq_31}); The values of $m_{1}$ at which NDTR can occur are all less than $1/(2\gamma)\approx-3.85$ (see inset for large values of $m_{o}$), which explains why NDTR cannot be observed when the value of either $T_{+}$ ($m_{1}=-1/2$) or $T_{-}$ ($m_{1}=1/2$) is fixed. The word 'possible' refers to the fact that any NDTR regime might be too narrow to be observed.
}\label{Fig_3}
\end{center}
\end{figure}

The conventional approaches of fixing the value of either $T_{+}$ or $T_{-}$, as described by Eqs. (\ref{Eq_14}) to \ref{Eq_16}), can be regarded as particular cases of a more general constraint
\begin{equation}\label{Eq_20}
\overline{T}=m_{o}+m_{1}\Delta T
\end{equation}
where $m_{o}\geq0$, $m_{1}\in(-\infty,+\infty)$ and
\begin{equation}\label{Eq_21}
T_{\pm}=m_{o}+(m_{1}\pm1/2)\Delta T
\end{equation}
The linear relation described by Eq. (\ref{Eq_20}) can be rearranged into the form of Eq. (\ref{Eq_11}). The condition
\begin{equation}\label{Eq_22}
T_{-}=m_{o}+(m_{1}-1/2)\Delta T\geq0
\end{equation}
implies
\begin{eqnarray}\label{Eq_23}
0\leq\Delta T\leq\Delta T_{max}=\frac{m_{o}}{1/2-m_{1}}
\end{eqnarray}
for $m_{1}<1/2$, and the absence of a $\Delta T_{max}$ for $m_{1}\geq1/2$. Using the relations
\begin{equation}\label{Eq_24}
n_{2}(\overline{T})=-\gamma/[1+t/\overline{T}]
\end{equation}
[from Eqs. (\ref{Eq_1}) and (\ref{Eq_9})] and
\begin{equation}\label{Eq_25}
n_{1}(\overline{T},\Delta T)=m_{1}\Delta T/\overline{T}
\end{equation}
[from Eqs. (\ref{Eq_13}) and (\ref{Eq_20})] as well as Eq. (\ref{Eq_19}), it can be shown that the occurrence of NDTR is conditioned upon a simultaneous satisfaction of the following inequalities:
\begin{equation}\label{Eq_26}
\Delta T>\Delta T^{*}\equiv\frac{m_{o}+t}{(\gamma-1)m_{1}}
\end{equation}
\begin{equation}\label{Eq_27}
(\gamma-1)m_{1}>0
\end{equation}
and
\begin{equation}\label{Eq_28}
\Delta T^{*}<\Delta T_{max}
\end{equation}
where the last inequality is needed only for cases of $m_{1}<1/2$. Using Eq. (\ref{Eq_23}) as well as Eqs. (\ref{Eq_26}) to (\ref{Eq_28}), one obtains the following conditions for the existence of a finite NDTR regime at $m_{1}<1/2$:
\begin{equation}\label{Eq_29}
m_{1}>\frac{1}{2}\left[\frac{1+(t/m_{o})}{\gamma+(t/m_{o})}\right]
\end{equation}
for $m_{1}>0$ and $\gamma>1$ (e.g. the $\phi^{4}$ model), and
\begin{equation}\label{Eq_30}
m_{1}<\frac{1}{2}\left[\frac{1+(t/m_{o})}{\gamma+(t/m_{o})}\right]
\end{equation}
with
\begin{equation}\label{Eq_31}
m_{o}>-t/\gamma
\end{equation}
for $m_{1}<0$ and $\gamma<0$ (e.g. the FPU-$\beta$ model). Fig. \ref{Fig_3} indicates graphically the theoretical combinations of $m_{1}$ and $m_{o}$ at which there exists a NDTR regime, for both the $\phi^{4}$ and the FPU-$\beta$ model (Fig. \ref{Fig_1}). It can be seen that, for the $\phi^{4}$ model ($\gamma\approx1.313\pm0.009$, $t \approx 0.52$ and $N = 32$), choosing $m_{1}=1/2$, i.e. fixing the value of $T_{-}$ at $m_{o}$, is just one of infinite possible ways of generating NDTR, and that choosing $m_{1}=-1/2$, i.e. fixing the value of $T_{+}$ at $m_{o}$, will not lead to an occurrence of NDTR. And for the FPU-$\beta$ model ($\gamma\approx-0.130\pm0.002$, $t \approx 3.73$ and $N = 64$), the conventional approaches of fixing the value of $T_{+}$ or $T_{-}$ will all prohibit the occurrence of NDTR, which requires $m_{1}$ to be no greater than $1/(2\gamma)\approx-3.85$ [obtained at $m_{o}\rightarrow\infty$ from Eq. (\ref{Eq_30})]. On the other hand, for $0<\gamma<1$, Eq. (\ref{Eq_27}) implies $m_{1}<0$, which contradicts Eq. ({\ref{Eq_29}}) of $m_{1}>0$ for a general case of $\gamma>0$; hence the overall absence of a NDTR regime. In addition, $\gamma=0$ implies $n_{2}(\overline{T})=0$, which violates Eq. (\ref{Eq_19}), and $\gamma=1$ implies $\Delta T^{*}\rightarrow\infty$, which violates Eq. (\ref{Eq_28}). All these cases without the presence of a NDTR regime implies that the constraint described by Eq. (\ref{Eq_20}) needs to be modified.

\begin{figure}[htbp]
\begin{center}\includegraphics[width=7.0125cm,height=5.41875cm]{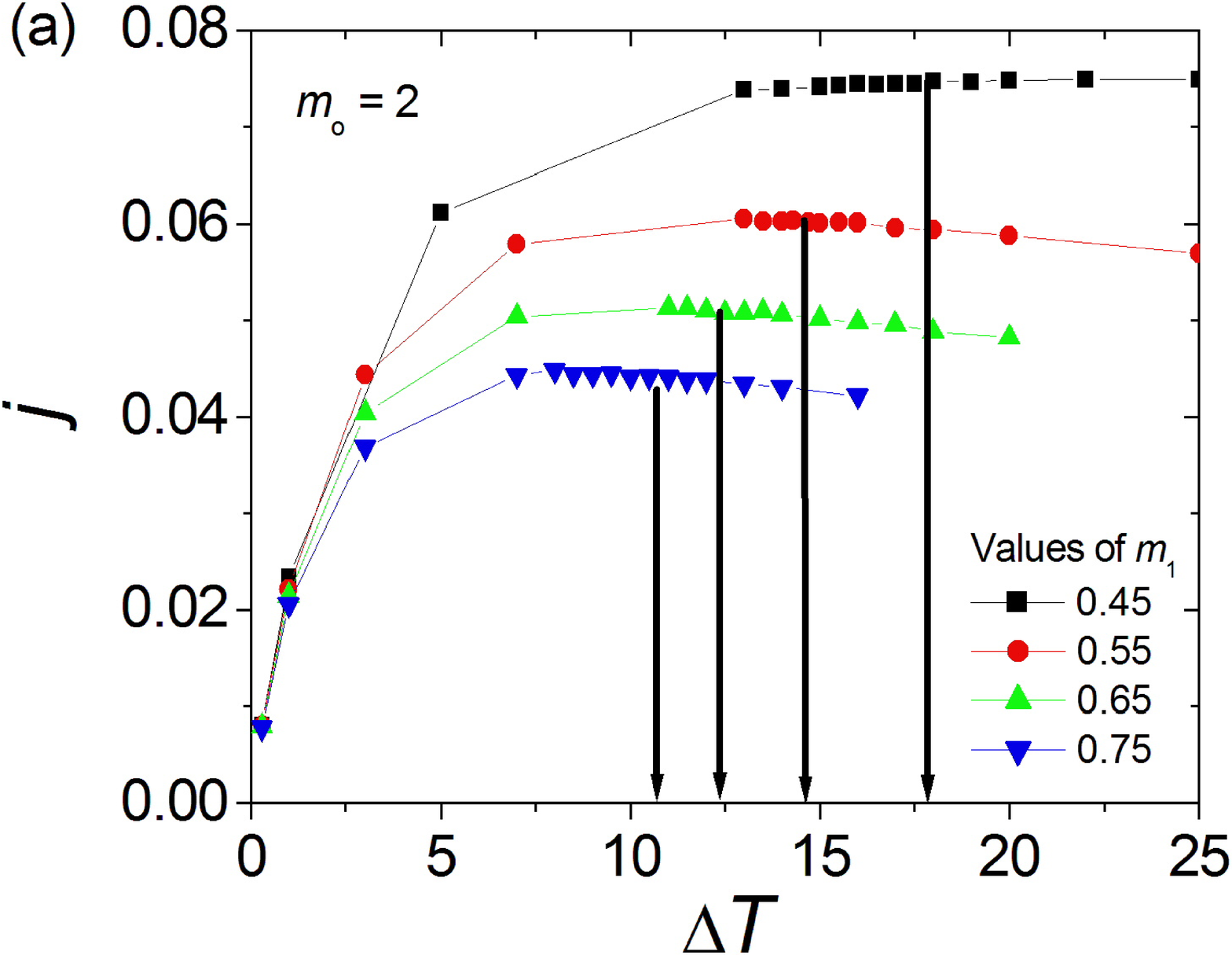}
\includegraphics[width=7.0125cm,height=5.41875cm]{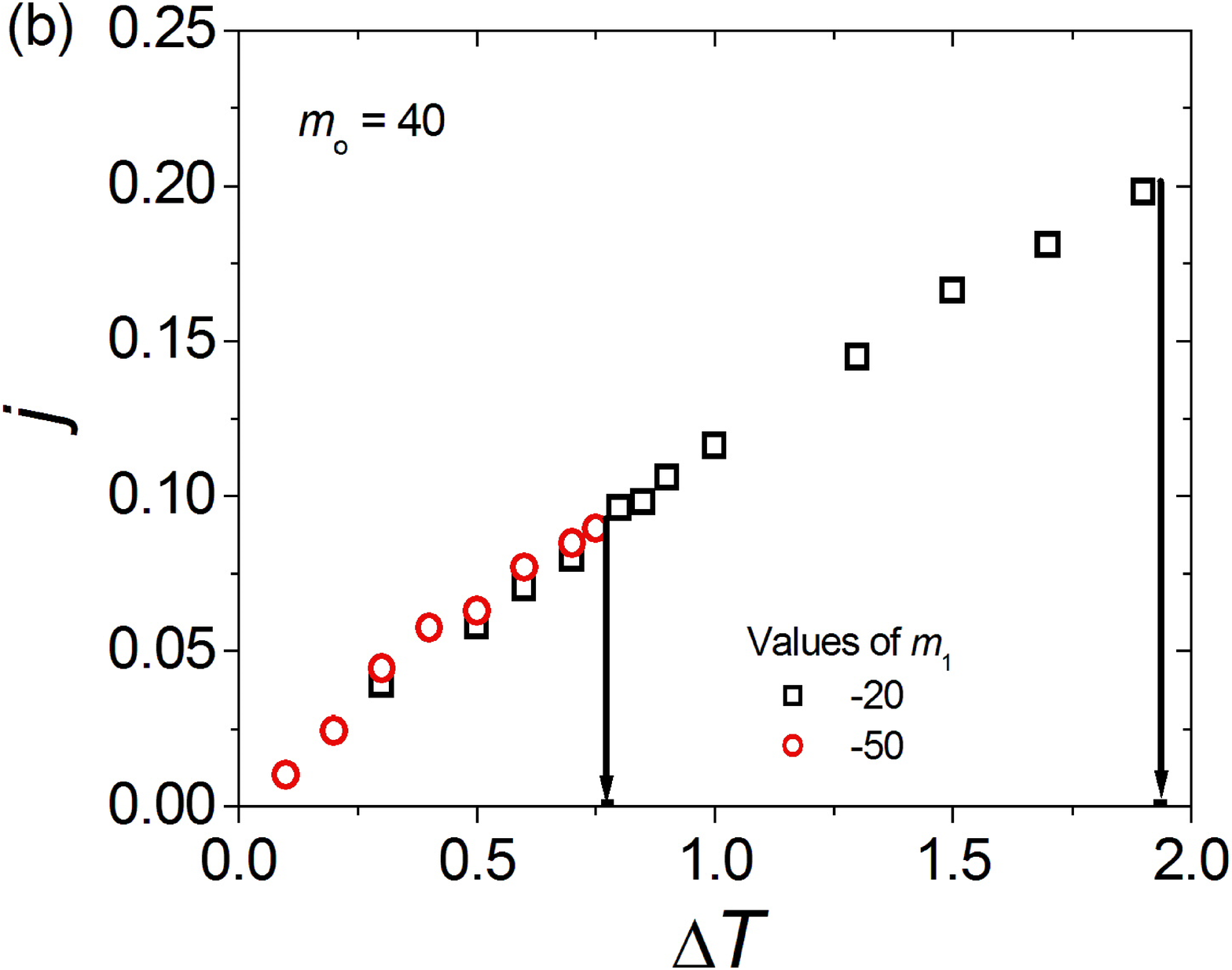}
\caption{Numerical results of heat flux $j$ as a function of the temperature difference $\Delta T$ at various combinations of $m_{o}$ and $m_{1}$, for (a) a $\phi^4$ chain of $N=32$ particles [$\gamma\approx1.313\pm0.009$ and $t \approx 0.52$] and (b) a FPU-$\beta$ chain of $N=64$ particles [$\gamma\approx-0.13\pm0.002$ and $t \approx 3.73$]. The vertical lines indicate the theoretical values of $\Delta T^{*}$ as predicted from Eq. (\ref{Eq_26}). For the FPU-$\beta$ model, the values of $\Delta T^{*}$ are practically equal to those of $\Delta T_{max}$ (Eq. (\ref{Eq_32})) so that NDTR cannot be observed.
}\label{Fig_4}
\end{center}
\end{figure}

The above theoretical predictions have been verified numerically: For the $\phi^{4}$ model, Fig. \ref{Fig_4}(a) shows a plot of the heat flux $j$ against the temperature difference $\Delta T$ for $m_{o}=2$ at various values of $m_{1}$. The transition from PDTR to NDTR for increasing $\Delta T$ occurs at values of $\Delta T^{*}$ that are very close to the predictions of Eq. (\ref{Eq_26}), where this equation suggests a $\Delta T^{*}\sim1/m_{1}$ dependence for $m_{1}>0$. The small discrepancies between theory and simulation are due to uncertainties in the fitted values of $\gamma$ and $t$. For the FPU-$\beta$ model, however, the relatively small magnitude of $\gamma<0$ suggests that $\Delta T^{*}$ would be so close to $\Delta T_{max}$ that the NDTR regime is too narrow to be observed (Fig. \ref{Fig_4}(b)):
\begin{equation}\label{Eq_32}
\frac{\Delta T^{*}}{\Delta   T_{max}}=\frac{\left(1+\frac{t}{m_{o}}\right)\left(\frac{1}{2m_{1}}-1\right)}{\left(\gamma-1\right)}\approx1
\end{equation}
since $m_{o}>>t$, $|m_{1}|>>1$ and $|\gamma|<<1$ [see Fig. \ref{Fig_4}(b) for some values of $m_{o}$ and $m_{1}$]. Such a weak dependence of $\kappa_{e}(\overline{T})$ on $\overline{T}$ implies that, for the constraint described by Eq. (\ref{Eq_20}), $n_{2}(\overline{T})$ is almost too small for Eq. (\ref{Eq_19}) to be satisfied. Yet, this can be overcome by resorting to other ways of achieving a sufficiently large magnitude of $n_{1}(\overline{T},\Delta T)$. As an example, consider a change of the boundary temperatures from $(T_{+},T_{-})=(51,49)$ to $(T_{+},T_{-})=(41.005,38.995)$. The value of $\overline{T}$ decreases from 50 to 40, while the value of $\Delta T$ increases from 2 to 2.01. According to Eq. (\ref{Eq_13}), the value of $n_{1}(\overline{T},\Delta T)$ for such a change is given by
\begin{equation}\label{Eq_33}
n_{1}(\overline{T},\Delta T)=\frac{\Delta T}{\overline{T}}\frac{d\overline{T}}{d\Delta T}\approx-44.56
\end{equation}
Using Eqs. (\ref{Eq_24}) and (\ref{Eq_33}), it can be shown that
\begin{equation}\label{Eq_34}
n_{1}(\overline{T},\Delta T)n_{2}(\overline{T}) \approx-5.35 <-1
\end{equation}
which matches the condition described by Eq. (\ref{Eq_19}) for the occurrence of NDTR. Such a prediction for the FPU-$\beta$ model has been verified numerically, where the heat flux $j$ was found to decrease from 0.26637 to 0.24569.

\section {Conclusions}

The numerical results presented here prove that NDTR can be generated not only by the conventional approaches of fixing the value of either $T_{+}$ or $T_{-}$, but also by a range of previously unexplored methods [i.e. constraints that can be expressed in the form of Eq. (\ref{Eq_11})] for varying the temperature difference $\Delta T$, where the linear relation described by Eq. (\ref{Eq_20}) represents only one of many different possibilities. For the purpose of generating NDTR, as long as the thermal conductivity varies with the system's average temperature, i.e. $n_{2}(\overline{T},\Delta T)\neq0$, one can always adjust the constraint $F(\overline{T},\Delta T)=0$ so as to produce a suitable value of $n_{1}(\overline{T},\Delta T)$ that satisfies Eq. ({\ref{Eq_18}}). As to future work, this theoretical analysis of NDTR, which is independent of specific temperature scales, will be useful when designing thermal devices based on components of one-dimensional heat flow \cite{R01,R02,R03}. It can also serve as a guide for manipulating heat flow in experimental systems, such as carbon nanotubes and boron-nitride nanotubes \cite{R18,R19}.

\section {Acknowledgements}

Stimulating discussions with members of the Centre for Nonlinear Studies, Hong Kong Baptist University, are grateful acknowledged. D. H. would like to acknowledge financial support from NSFC (Grants 11047185 and 11105112) and FRFCU (Grant 2010121009) of the People's Republic of China.

\end{document}